\documentstyle[aps,aps12,epsf]{revtex}

\def\ham{{\bf \rm H}}

\def\bmat{{\bf \rm B}}
\def\cmat{{\bf \rm C}}

\def\spose#1{\hbox to 0pt{#1\hss}}
\def\ltapprox{\mathrel{\spose{\lower 3pt\hbox{$\mathchar"218$}}
 \raise 2.0pt\hbox{$\mathchar"13C$}}}
\def\gtapprox{\mathrel{\spose{\lower 3pt\hbox{$\mathchar"218$}}
 \raise 2.0pt\hbox{$\mathchar"13E$}}}

\begin{document}

%\draft

\title{\vspace*{-3cm}\hfill\mbox{\small FSU-SCRI-99-04}\\
\vspace*{1cm}Approach to the Continuum Limit of the Quenched Hermitian 
Wilson--Dirac Operator}
\author{
Robert G. Edwards, Urs M. Heller, Rajamani Narayanan}
\address{
SCRI, Florida State University, 
Tallahassee, FL 32306-4130, USA}

\maketitle 

\begin{abstract}

We investigate the approach to the continuum limit of the spectrum of
the Hermitian
Wilson--Dirac operator in the supercritical mass region for pure
gauge SU(2) and SU(3) backgrounds.  For this we study the spectral flow of the
Hermitian Wilson--Dirac operator in the range $0\le m\le 2$.  We find
that the spectrum has a gap for $0 < m\le m_1$ and that the spectral
density at zero, $\rho(0;m)$, is non-zero for $m_1\le m\le 2$. 
We find that $m_1\rightarrow 0$ and, for $m \ne 0$, $\rho(0;m)\rightarrow 0$ 
(exponential in the lattice spacing)
as one goes to the continuum limit. We also compute the
topological susceptibility and the size distribution of the zero
modes. The topological susceptibility scales well in the lattice
spacing for both SU(2) and SU(3).  The size distribution of the zero
modes does not appear to show a peak at a physical
scale.

\end{abstract}


\section{Introduction}

Continuum gauge field theory works under the assumption that all
fields are smooth functions of space--time. This assumption is
certainly a valid one for quantum gauge field theories that respect
gauge invariance: One should always be able to fix a gauge so that the
gauge fields are smooth functions of space--time since the action that
contains derivatives in gauge fields will not allow it otherwise. The
space of smooth gauge fields typically has an infinite number of
disconnected pieces where the number of pieces is in one to one
correspondence with the set of integers~\cite{topology}.
Every gauge field in each
piece can be smoothly interpolated to another gauge field in the same
piece but there is no smooth interpolation between gauge fields in
different pieces.  This is the case for U(1) gauge fields in two
dimensions and SU(N) gauge fields in four dimensions.

In lattice gauge theory, gauge fields are represented by link variables
$U_\mu(x)$ 
that are elements of the gauge group. Continuum derivatives are 
replaced by finite differences and the concept of smoothness of
gauge fields does not apply.
Any lattice gauge field configuration, $U_\mu(x)=e^{iA_\mu(x)}$
can be deformed to the trivial gauge field configuration by the
interpolation $U_\mu(x;\tau)=e^{i\tau A_\mu(x)}$ with 
$U_\mu(x;1)=U_\mu(x)$ and $U_\mu(x;0)=1$. Since smoothness does not
hold on the lattice away from the continuum limit, the space of gauge
fields on the lattice forms a simply connected space. Separation of the
gauge field space into an infinite number of disconnected pieces can only
be realized in the continuum limit. 

In this paper, we will address the following basic question: Do we see
a separation of lattice gauge fields configurations into topological
classes as we approach the continuum limit? To answer this question,
we will use several ensembles of lattice gauge field configurations
obtained from pure SU(2) and SU(3) gauge field theory.
We will use a Wilson--Dirac fermion
to {\sl probe} the lattice gauge field configuration. 
Our motivation is
the overlap formalism~\cite{over} for chiral gauge theories. 
Topological aspects
of the background gauge field are properly realized by the chiral fermions
in this formalism and therefore it provides a good framework to answer
the above question. 
The hermitian Wilson--Dirac operator enters 
the construction of lattice chiral fermions in the overlap formalism
and topological properties of the gauge fields are studied by looking
at the spectral flow of the hermitian Wilson--Dirac operator as a 
function of the fermion mass. 
Contrary to some other approaches to investigate the topological
properties of lattice gauge field configurations~\cite{topol_rev}
we do not modify the gauge fields, generate by some Monte Carlo
procedure, in any way.

The paper is organized as follows. We begin by explaining in 
Section~\ref{sec:specflow} the connection
between the spectral flow of the hermitian Wilson--Dirac operator and
the topological content of the background gauge field. 
Possible scenarios for the qualitative nature of the spectrum on the
lattice are presented.
In Section~\ref{sec:rho0} we present numerical results on the spectral
properties of lattice gauge field ensembles and their behavior as
the continuum limit is approached. Results for the topological
susceptibility in pure SU(2) and SU(3) gauge theory computed using the 
overlap definition of the topological charge are shown in 
Section~\ref{sec:suscep}. We also present
results on the size distribution of the zero modes of the
Hermitian Wilson--Dirac operator.

\section{Spectral flow, topology and condensates}
\label{sec:specflow}

\begin{figure}
\epsfxsize=5in
\centerline{\epsfbox[100 175 500 500]{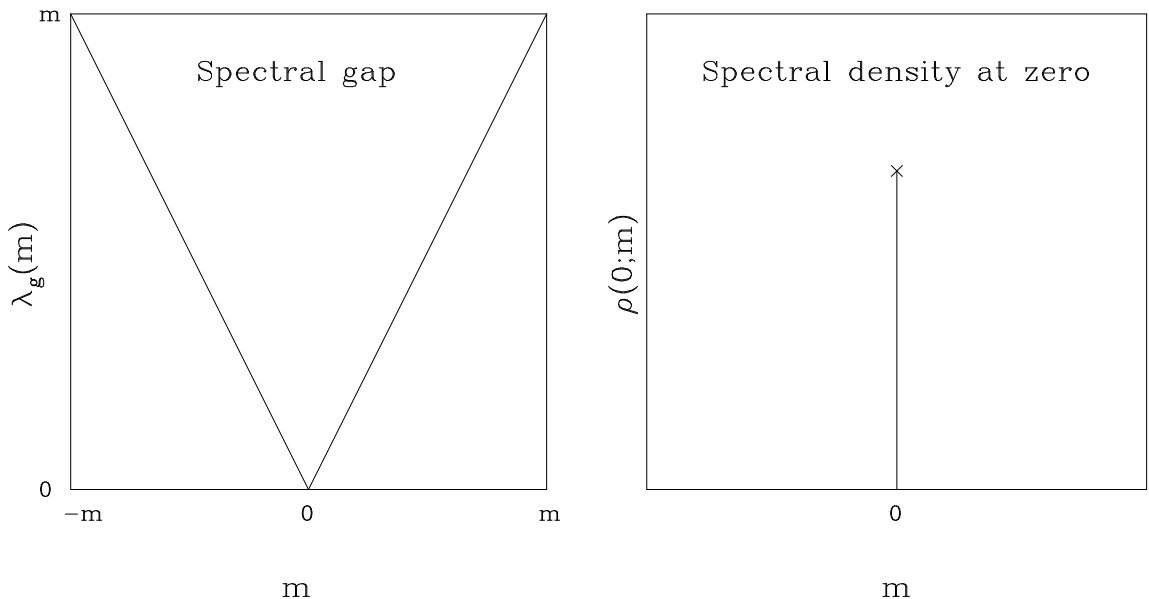}}
\caption{Continuum picture of the spectral gap and the spectral
density at zero.}
\label{fig:cont}
\end{figure}

The massless Dirac operator in the continuum anticommutes with
$\gamma_5$. Therefore, the non-zero imaginary eigenvalues of
the massless Dirac operator come in pairs, $\pm i\lambda$, with
$\psi$ and $\gamma_5\psi$ being the two eigenvectors.
The zero eigenvalues of the massless Dirac operator are also
eigenvalues of $\gamma_5$. These chiral zero modes are a 
consequence of the topology of the background gauge field.
It is useful to consider the spectral flow of the Hermitian
Dirac operator:
\begin{equation}
\ham(m) = \gamma_5 (\gamma_\mu D_\mu - m )
\end{equation}
The non-zero eigenvalues of the massless Dirac operator
combine in pairs to give the following eigenvalue equation:
\begin{equation}
\ham(m) \chi_\pm = \lambda_\pm(m) \chi_\pm = 
 \pm \sqrt{\lambda^2+m^2} \chi_\pm ~.
\end{equation}
$\chi_\pm$ are linear combinations of $\psi$ and $\gamma_5\psi$.
The eigenvalues $\lambda_\pm(m)$ of
these modes never cross the x-axis in the spectral flow
of $\ham(m)$ as a function of $m$. 
The zero eigenvalues, $\gamma_\mu D_\mu\phi_\pm =0$ with
$ \gamma_5 \phi_\pm = \pm \phi_\pm$ result in
\begin{equation}
\ham(m) \phi_\pm = \mp m \phi_\pm
\end{equation}
These modes, associated with topology, result in flow lines
that cross the x-axis. A positive slope corresponds to negative
chirality and vice-versa. The net number of lines crossing zero
(the difference of positive and negative crossings) is the
topology of the background gauge field. 
Global topology of gauge fields cause exact zero eigenvalues at
$m=0$. In addition, one can have a non-zero spectral
density at zero. 
In an infinite volume in the continuum, the spectrum is continuous and
$\rho(\lambda;m)d\lambda$ is the number of eigenvalues in the
infinitesimal region $d\lambda$ around $\lambda$. The spectral gap $\lambda_g(m)$
defined as the lowest eigenvalue at $m$
is equal to $|m|$. The spectral density at zero, $\rho(0;m)$, can be
non-zero only at $m=0$ indicating spontaneous chiral symmetry breaking
in a theory like QCD. The continuum picture is shown in Fig.~\ref{fig:cont}.

To study the possible emergence of the above picture 
as the continuum limit of a lattice gauge theory picture, we need
to have a lattice realization of $\ham(m)$. It is important to
note that we are interested in the spectral flow of a single
Dirac fermion. With this in mind, we choose the hermitian
Wilson--Dirac operator obtained by multiplying the standard Wilson--Dirac
operator by $\gamma_5$:
\begin{equation}
\ham_L(m) = 
\pmatrix {\bmat(U) -m & \cmat(U) \cr \cmat^\dagger(U) &  -\bmat(U) +m \cr}.
\end{equation}
$\cmat$ is the naive lattice first derivative term
and $\bmat$ is the Wilson term.
We are interested in the spectral flow of $\ham_L(m)$ as a function
of $m$. We note that 
$m=0,2,4,6,8$ are the points where the free fermions become massless
with degeneracies $1,4,6,4,1$ respectively. 
Next we observe that 
$\ham_L(m)$ can have a zero eigenvalue only if $m > 0$~\cite{over1}.

\begin{figure}
\epsfysize=5.5in
\centerline{\epsfbox[150 100 500 550]{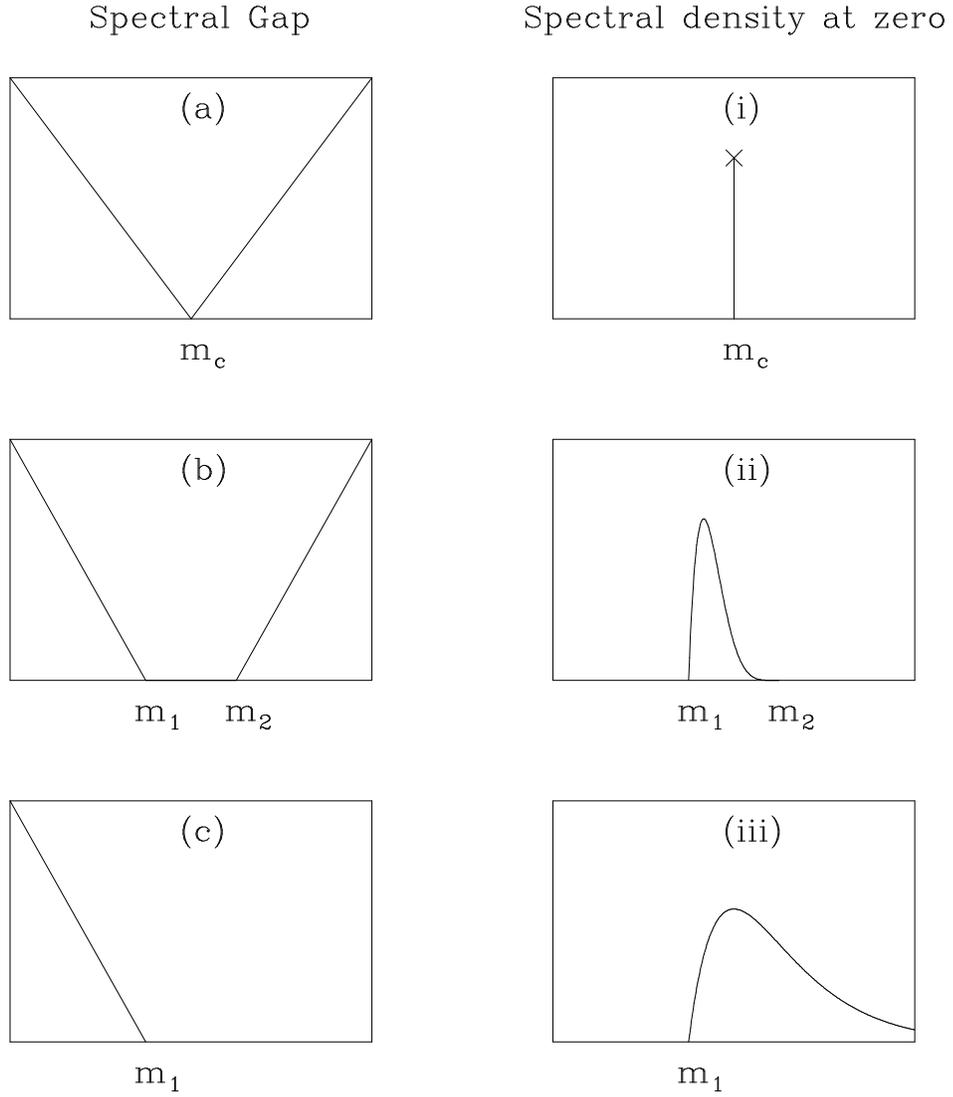}}
\caption{Possible scenarios of the spectral gap and the spectral
density at zero on the lattice.}
\label{fig:latt}
\end{figure}
We focus on the range $0\le m \le 2$ and propose the following
scenarios for the spectral gap and the spectral density at zero
on the lattice and their approach to the continuum limit.
Six different, but not completely independent
scenarios are possible as shown in Fig.~\ref{fig:latt}.
\begin{itemize}
\item On the lattice we have (a) and (i) with $m_c\rightarrow 0$
in the continuum limit. $\rho(0;m_c)$ approaches the continuum
limit with proper scaling taken into account.
\item On the lattice we have (b) and (ii) where (a) and (i) are the
continuum limit. In this case,  $m_1\rightarrow 0$ and
$m_2\rightarrow 0$. In the limit we also get $\rho(0;0)$.
\item On the lattice we first have (c) and (ii) going, at weaker coupling
to (b) and (ii), where (a) and (i) are the continuum limit.
The gap opens up at some $m_2 > m_1$ at some coupling and afterwards
the approach to the continuum is as in the previous scenario.
\item On the lattice we have (c) and (ii) where (c) and (i) are the
continuum limit. In this case,  $m_1\rightarrow 0$. However, the gap 
does not open up for $m > 0$ in the continuum limit.
\item On the lattice we first have (c) and (iii), going to (b) and (ii)
at some coupling. Afterwards, the approach to the continuum limit is
again as in the second scenario.
\item {\it On the lattice we have (c) and (iii) where (c) and (i) are the
continuum limit. Here also  $m_1\rightarrow 0$
and $\rho(0;m)=0$ if $m > 0$. However, the gap does not open up
for $m > 0$.}
\end{itemize}

We will show that numerical studies of the spectral flow on various
ensembles favor the last scenario. Before we do that, 
we present a topological argument which will show that
zero eigenvalues of $\ham_L(m)$ can occur anywhere in
the region $0 < m < 8$~\cite{smooth}. 
The spectrum of $\ham_L(m)$ and $-\ham_L(8-m)$ are identical for
an arbitrary gauge field background. 
Since zero eigenvalues can occur only for $m> 0$ in
$\ham_L(m)$, it follows that zero eigenvalues can occur only in
the region $0 < m < 8$. It also follows that
every level crossing zero from above in the spectral flow of
$\ham_L(m)$ must be accompanied by a level crossing zero from below.
In a single instanton background a level crossing zero from above
at $m_+$ is accompanied by another level crossing zero from below
at $2 > m_- > m_+$. The second crossing is due to
one of the four doubler modes.
Both $m_\pm$ will be functions of the size
of the instanton $\rho$ in lattice units. For $\rho >> a$, 
$m_+ \approx 0$ and $m_- \approx 2$. As $\rho$ decreases,
$m_+$ moves farther away from zero and $m_-$ moves away from
$2$ and closer to $m_+$. This motion as a function of $\rho$
is smooth and for some value of $\rho$, $m_+=m_-$.
The spectral flow changes smoothly as the configuration is
changed slowly. As we move in configuration space the topological
charge of a configuration changes. Tracing the spectral flow as
a function of configurations shows that
zero eigenvalues of $\ham_L(m)$ can occur anywhere in
the region $0 < m < 8$.

\section{Spectral density at zero}
\label{sec:rho0}

In the previous section, we argued that $\ham_L(m)$ can have zero
crossings anywhere in the region $m_1 \le m \le 2$. Therefore the spectral
gap is zero in this region on the lattice. This has direct implications
for how the spectral density at zero behaves on the lattice.
A careful study of the spectral density at zero has been performed
on a variety of SU(3) pure gauge ensembles. 
This is done by computing
the low lying eigenvalues of $\ham_L(m)$ using the Ritz functional~\cite{Ritz}.
The low lying eigenvalues over the whole ensemble are then used to
obtain the integral of the spectral density function,
$\int_0^\lambda \rho(\lambda^\prime;m)d\lambda^\prime$.
A linear fit in $\lambda$ is made, and $\rho(0;m)$ is obtained as the 
coefficient of the linear term. 

\begin{figure}
\epsfxsize=5in
%\centerline{\epsfbox[50 40 600 600]{rho0_talk.ps}}
\centerline{\epsffile{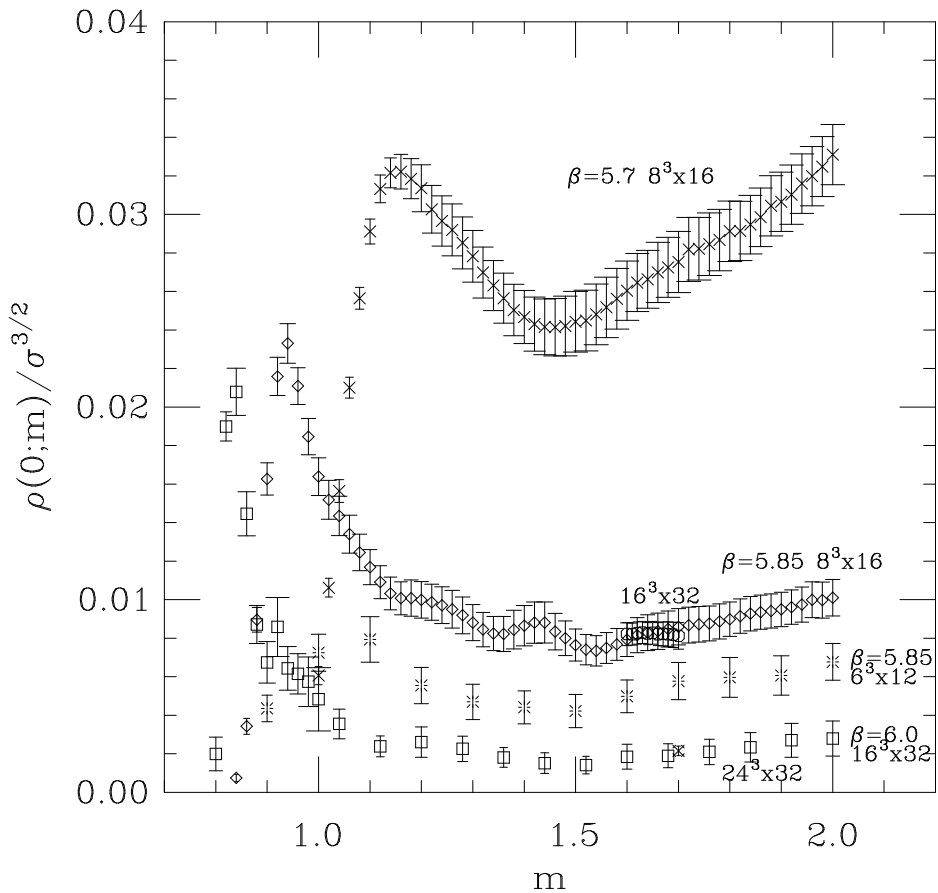}}
\caption{$\rho(0;m)$ as a function of $m$ for various SU(3) pure
gauge ensembles at gauge couplings $\beta=6.0$, $5.85$ and $5.7$.}
\label{fig:rho0_talk}
\end{figure}

All ensembles show a peak in $\rho(0;m)$ at some value of $m$ near
$m_1$. There is a sharp rise to the peak from the left and a gradual
fall from the peak on the right. There is a gradual rise again to a
second peak at the location of the first set of doublers.  The peak
itself gets sharper and moves to the left as one goes toward the
continuum limit.  $\rho(0;m)$ is non-zero for $m_1 \le m \le 2$ in the
infinite volume limit at any finite value of the lattice gauge
coupling (see below). $m_1$ goes to zero as the lattice coupling
approaches the continuum. $\rho(0;m)$ approaches the infinite lattice
volume limit from below as expected. We are fairly confident that we
have the infinite volume limit estimate for $\rho(0;m)$ at all the
lattice spacings plotted in Fig.~\ref{fig:rho0_talk}.

\begin{figure}
\epsfxsize=3.5in
%\centerline{\epsfbox[100 30 500 600]{rho0_scale_m1p7.ps}}
\centerline{\epsffile{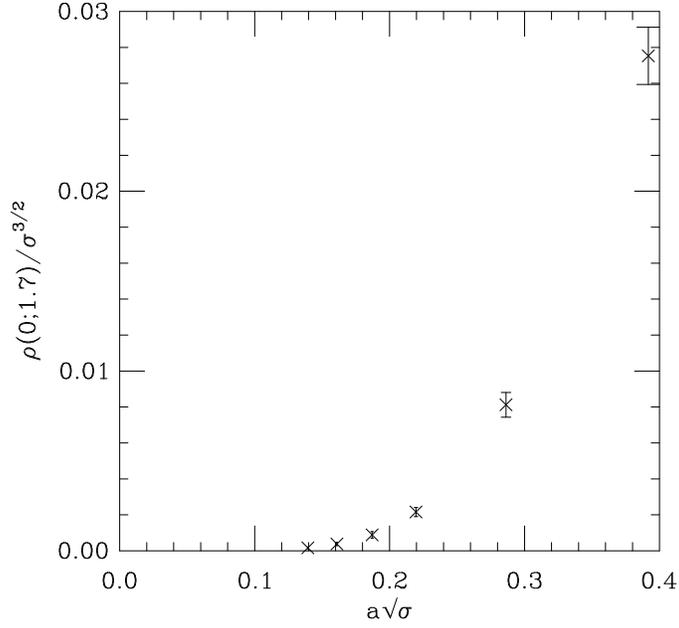}}
\caption{The approach of $\rho(0;1.7)$ to the continuum limit as a
function of the lattice spacing in units of the
string tension for $\beta=5.7$, $5.85$, $6.0$, $6.1$, $6.2$ and $6.3$.}
\label{fig:rho0_scale}
\end{figure}

\begin{figure}
\epsfxsize=3.5in
%\centerline{\epsfbox[100 30 500 600]{rho0_scale_exp_m1p2-1p7.ps}}
\centerline{\epsffile{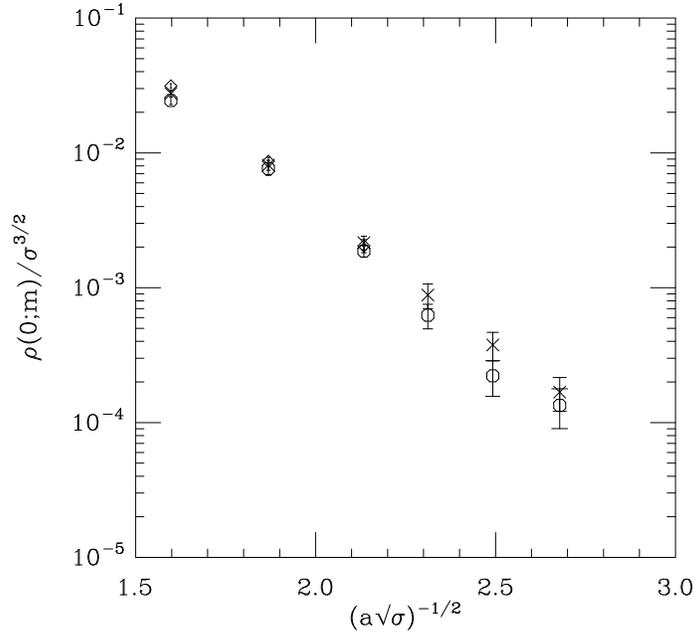}}
\caption{The approach of $\rho(0;m)$ to the continuum limit as a
function of $1/\sqrt{a\sqrt{\sigma}}$ for $m=1.2$ (diamonds) $1.5$ 
(octagons) and $1.7$ (crosses). 
For $m=1.2$ only values at $\beta\le 6.0$ are shown.}
\label{fig:rho0_scale_exp_all}
\end{figure}

In Fig.~\ref{fig:rho0_scale} and Fig.~\ref{fig:rho0_scale_exp_all}, we
focus on the behavior of $\rho(0;m)$ at a fixed $m$ as one approaches
the continuum limit.  
In Figure~\ref{fig:rho0_scale} we plot $\rho(0;m)$ 
as a function of the lattice spacing measured in units of the
square root of the string tension (the values for the string tension
are taken from Ref.~\cite{string}). In this figure
$\rho(0;m)$ appears to go to
zero exponentially in the inverse lattice spacing.
This is given some
credence by plotting the same figure in a logarithmic scale in
Fig.~\ref{fig:rho0_scale_exp_all} where the data is shown for several
values of $m$. For $\beta=5.7$, the peak in $\rho(0;m)$ is quite close
to $m=1.2$ as can be seen in Fig.~\ref{fig:rho0_talk}, resulting in
a large value for $\rho(0;1.2)$.

We remark that the $\rho(0;m)$ plotted in
Fig.~\ref{fig:rho0_scale_exp_all} seem to favor a functional form
fitting $be^{-c/\sqrt{a}}$ for each $m$. The power of $a$ in the
exponent is a consequence of an empirical fit but the data presents
substantial evidence for the following: $\rho(0;m)$ in the
supercritical mass region is non-zero for all finite lattice
spacings. The approach to zero at zero lattice spacing is faster than
any power of the lattice spacing. This shows that the last scenario
presented in the previous section is favored by our numerical results.

\section{Topological susceptibility}
\label{sec:suscep}

In addition to studying $\rho(0;m)$, we also looked at the
density of levels crossing zero in an infinitesimal range $dm$ centered
at $m$. In the continuum we expect levels crossing zero only at $m=0$, but
on the lattice we find a finite density of levels crossing zero wherever
$\rho(0;m)$ is non-zero on the lattice.
The overlap formalism for constructing a chiral gauge theory on the
lattice~\cite{over} provides a natural definition of the index, $I$,
of the associated chiral Dirac operator. The index is equal to half
the difference of negative and positive eigenvalues of the hermitian
Wilson--Dirac operator.
 A simple way to compute the index $I$
is to compute the lowest eigenvalues of $\ham_L(m)$ at some suitably
small $m$ before any crossings of zero occurred.
Then $m$ is slowly varied and the number and direction of zero
crossings are tracked.  The net number at some $m_t$ is the index of
the overlap chiral Dirac operator.
Since crossings occur for all values of $m$ in the range $m_1 \le m \le 2$,
the topological
charge of a lattice gauge field configuration defined as the net
level crossings in $\ham_L(m)$ in the range $[0,m_t]$ will depend on
$m_t$.

\begin{figure}
\epsfxsize = 5in
\centerline{\epsfbox[50 40 600 600]{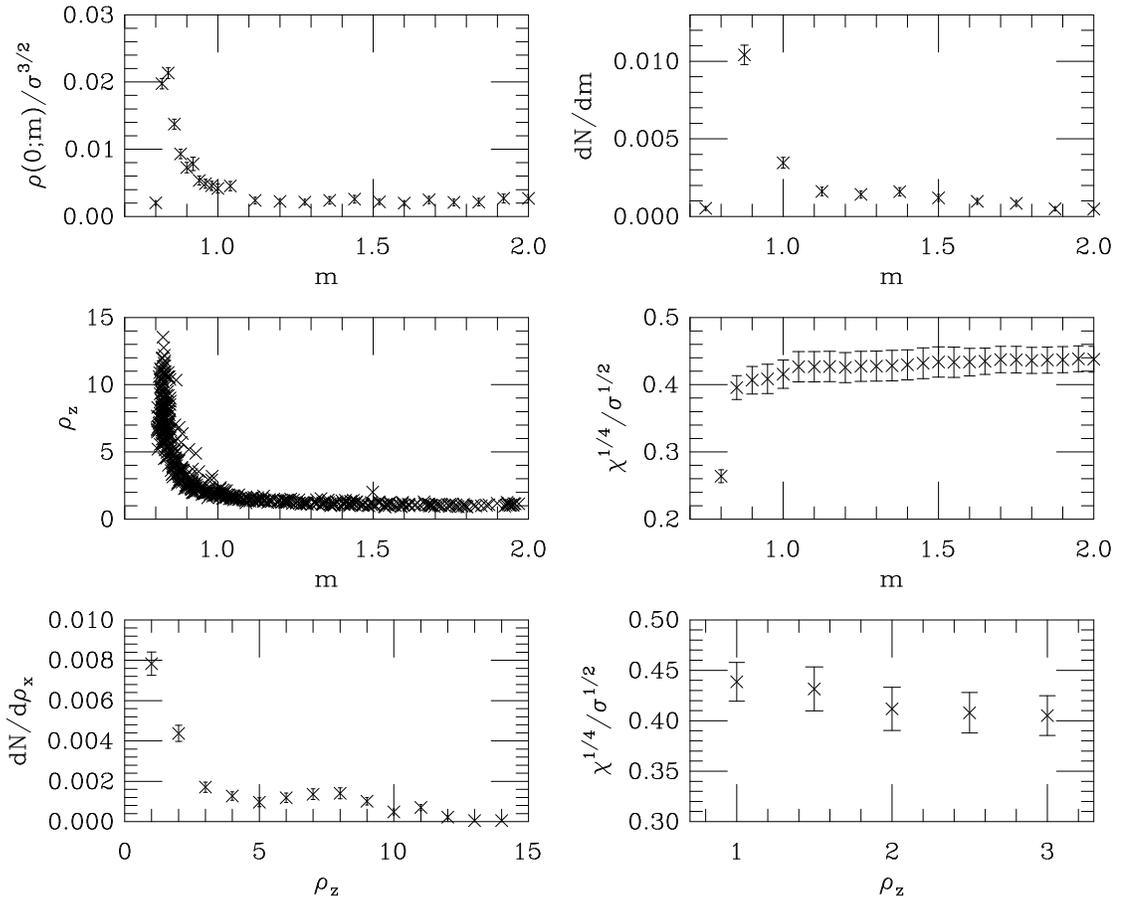}}
\caption{
  Detailed study for $\beta=6.0$, $16^3\times 32$.
}
\label{b6.0_all}
\end{figure}

The topology of a single lattice gauge field
configuration is not interesting in a field theoretic sense. One has
to obtain an ensemble average of the topological susceptibility and
study its dependence on $m$. This has been done on a variety of
ensembles
and the results show that the topological susceptibility is
essentially independent of $m$ in the region to the left of the peak
in $\rho(0;m)$. A detailed study of the SU(3) ensemble at $\beta=6.0$
on a $16^3\times 32$ lattice 
presented in Figure~\ref{b6.0_all} illustrates this point.
In the first line is shown the
density of zero eigenvalues $\rho(0;m)$ and the number of crossings in
each mass bin. 
%Since there are a nonzero number of crossings, we see
%that $\rho(0;m)$ does indeed measure zero eigenvalues, and not just
%small eigenvalues near zero. 
We see that $\rho(0;m)$ rises
sharply in $m$, then falls to a nonzero value where there is a small
number of levels crossing zero.
In the second line of Figure~\ref{b6.0_all}, we show the size of the
zero modes $\rho_z(m)$. 
We define a size of the eigenvector associated with the level crossing
zero mode as
\begin{displaymath}
\rho_z(m) = {\frac{1}{2}}  {{\sum_t f(t)} \over {f_{\rm max}}}
\quad
%\end{equation}
%\begin{equation}
f(t) = 
\sum_{\vec x} {\rm tr}(\phi^\dag({\vec x},t)\phi({\vec x},t))
\nonumber
\end{displaymath}
%motivated by the t'Hooft zero mode \ $\rho_z^2 / 2(t^2 + \rho_z^2)^{3/2}$
% 
where $\phi(\vec x,t)$ is the eigenvector of $\ham_L$
at the crossing point and $f_{max}$ is the maximum value of $f(t)$
over $t$. Another definition
based on the second moment of $f(t)$
was used in Ref.~\cite{su3_top}. We should emphasize
that we look only at the sizes of eigenmodes that cross, and only
close to the crossing point. Only then can we expect to get a good
estimate of the localization size inspired by the 't Hooft zero mode.
The modes are large near $m_1$ where
$\rho(0;m)$ is large, then $\rho_z(m)$ drops sharply to about $1$ or $2$ lattice
spacings and stays there up to $m=2$. We see that the corresponding 
topological susceptibility 
rises sharply when $\rho_z(m)$ is large for $m$ near
$m_1$ and then it is quite stable when $\rho_z(m)$ is small. This result shows
that while the index, $I$, of the field is $m$ dependent,
the topological susceptibility, $\chi$  
(a physical quantity) is independent of
the contribution from the small modes for $m \gtapprox 1$. 

To further clarify the relative contribution of the zero modes, in the
last line of Figure~\ref{b6.0_all} the zero mode size distribution is
plotted as a function of $\rho_z$. 
In the adjacent graph, the topological susceptibility,
$\chi$, here defined by the contribution of zero modes of size $\rho_z$ and
larger, is stable when $\rho_z < 2$. Hence, the small modes do not
affect the estimate of $\chi$ even though there is an abundance of such
modes. Our estimates of $\chi$ are shown in
Table~\ref{tab:results} where we use the string tension value
$\sqrt\sigma = 440$ MeV to set the scale. Our results are in rough
agreement with other groups~\cite{topol} and also show good
evidence for scaling.

\begin{table}
\caption{Topological susceptibility and parameters
for $SU(3)$ and $SU(2)$.}
\label{tab:results}
\begin{tabular}{|c||l|c|r|l|}
\hline
&$\beta$ & size & $N_{\rm conf}$ & $\chi^{1/4}$(MeV) \\
\hline
&$6.0$ & $16^3\times 32$ & 75 & 194(10) \\
$SU(3)$ &$5.85$ & $8^3\times 16$ & 200 & 198(05) \\
&$5.7$ & $8^3\times 16$ & 50 & 193(10) \\
\hline
&$2.6$ & $16^4$ & 400 & 229(05) \\
$SU(2)$ &$2.5$ & $16^4$ & 100 & 232(10) \\
&$2.4$ & $16^4$ & 200 & 220(06) \\
\hline
\end{tabular}
\end{table}

\section{Size distribution of zero modes}
\label{sec:size}

\begin{figure}
\epsfxsize = 5in
%\centerline{\epsfbox[50 40 600 600]{dN_su2.ps}}
\centerline{\epsffile{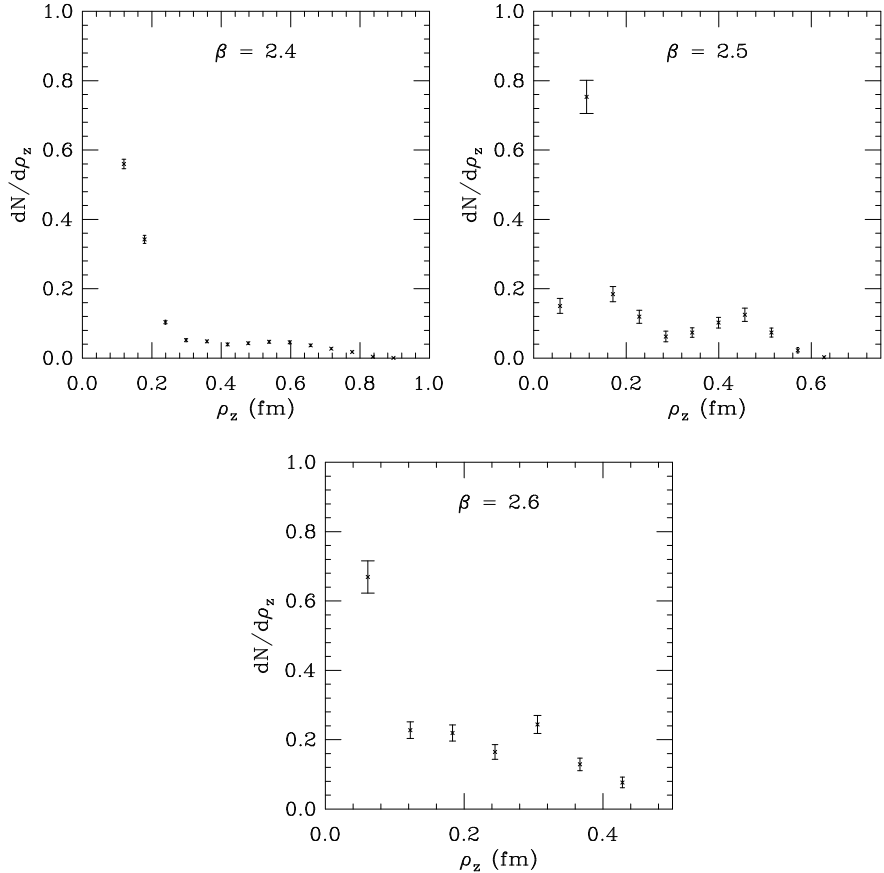}}
\caption{
 Size distribution of zero modes on various SU(2) ensembles.
}
\label{dN_su2}
\end{figure}

\begin{figure}
\epsfxsize = 4in
%\centerline{\epsfbox[50 40 600 600]{dN_su2.ps}}
\centerline{\epsffile{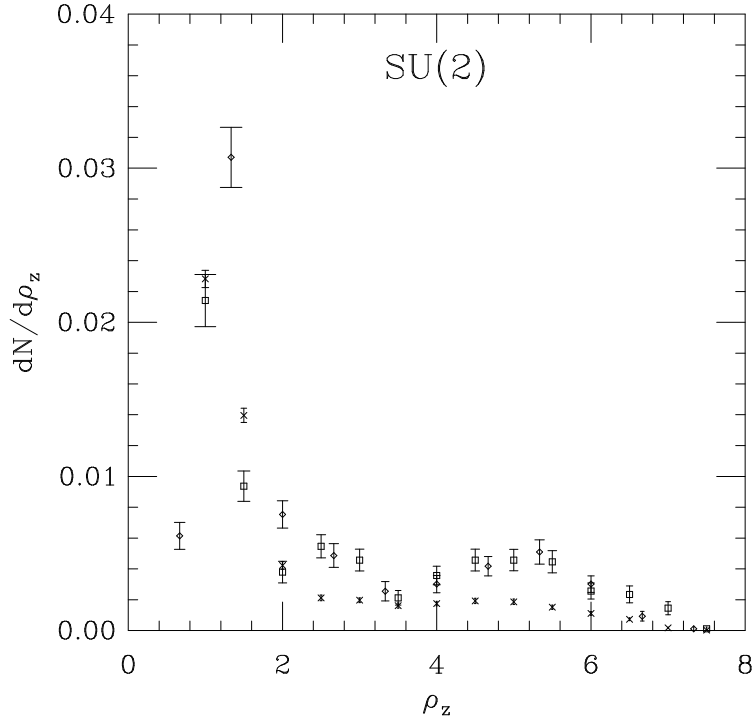}}
\caption{
Size distribution of zero modes on the SU(2) ensembles plotted in
lattice units. The crosses correspond to $\beta=2.4$, the diamonds
$\beta=2.5$, and the squares $\beta=2.6$. All lattice sizes are
$16^4$.
}
\label{dN_su2_rho}
\end{figure}

%In the previous section we showed that the pure gauge ensembles for which
%we studied the spectral flow gave good continuum estimates for the topological
%susceptibility. Our analysis also
%provides us with a distribution of the sizes of the zero modes
%as shown in Figure~\ref{b6.0_all}.
Studies in smooth gauge field backgrounds on the lattice have shown that
single instantons result in a single level crossing zero at some $m$ in
the region $[0,2]$ where the shape of the mode at the crossing is
a good representation of the shape of the instanton~\cite{smooth}. Similarly,
there is a pair of levels crossing zero (one from above and another
from below) when the gauge field background has an instanton and
anti-instanton. Again, the shape of the modes at the crossing points
are good representations of the instanton and anti-instanton. 
This motivates us to look at the size distribution of the zero
modes of the Hermitian Wilson-Dirac operator on the lattice.
%Since
%the gauge fields on the lattice are rough we do not expect the shape
%of the zero modes to truly reproduce the size and location content
%of topological objects in the gauge field background. But levels
%crossing zero affect the global topology of the background gauge field
Since lattice gauge fields generated in typical Monte Carlo simulations
are rough, the correspondence between zero modes and topological
objects might be questionable -- the existence of topological objects
(a collection of instantons and anti-instantons) ``underneath'' the
typically large quantum fluctuations is somewhat questionable as well.
However, levels crossing zero contribute to the global topology
and an analysis of the size distribution of the zero modes is therefore
interesting. Such size distributions are shown in Figure~\ref{dN_su2}
for gauge group SU(2).
All distributions show a sharp rise at small
sizes due to the abundance of small zero modes that occur in the
bulk of the region $m_1 \le m \le 2$. These modes do not affect the
computation of the topological susceptibility and can be viewed as being due
to the ultra-violet
fluctuations in the gauge field background. If we eliminate the small
modes from the distribution, the size distribution at $\beta=2.4$ on
the $16^4$ lattice shows some evidence for a broad peak around 
$\rho_z=0.6$ fm. One should keep in mind that the box size is roughly
$1.92$ fm and the peak is occurring at a value which is roughly a third
of the box. It is tempting to explain this peak as a finite volume effect.
Some support for this explanation is provided by looking at the distributions
at $\beta=2.5$ and $\beta=2.6$ on a $16^4$ lattice. These boxes are
now roughly $1.38$ fm and $0.98$ fm, respectively. After discarding the
small zero modes, both the distributions show a broad peak at roughly
$\rho_z=0.45$ fm and $\rho_z=0.3$ fm, respectively. 
As in the $\beta=2.4$ case these
peaks occur at roughly a third of the box size and the magnitude of
the peak is larger as one goes to weaker coupling for a fixed lattice
volume. This is quite consistent with the peak being a finite volume
effect. In Fig.~\ref{dN_su2_rho} we show all the SU(2) size
distributions together plotted in lattice units. There is evidence for
a broad peak at roughly $5$ lattice units -- roughly a third of the
lattice box size.
Therefore, we conclude that the size distribution of zero modes does not
show evidence for a peak at a physical scale 
even after we remove the small modes which are most likely lattice
artifacts.
%Due to the abundance of the small modes, the gauge field is clearly
%not smooth and cannot be thought of as being made up of instantons
%and anti-instantons. As such it is not possible to relate the size
%analysis here to any size distribution of topological objects
%and provide some input for the instanton liquid model~\cite{Shuryak}.
We have to conclude that it is not possible to relate the size
analysis of the zero modes carried out here to a size distribution
of topological objects as it is postulated for the instanton liquid
model of QCD~\cite{Shuryak}.

\begin{figure}
\epsfxsize = 5in
%\centerline{\epsfbox[50 40 600 600]{dN_su3.ps}}
\centerline{\epsffile{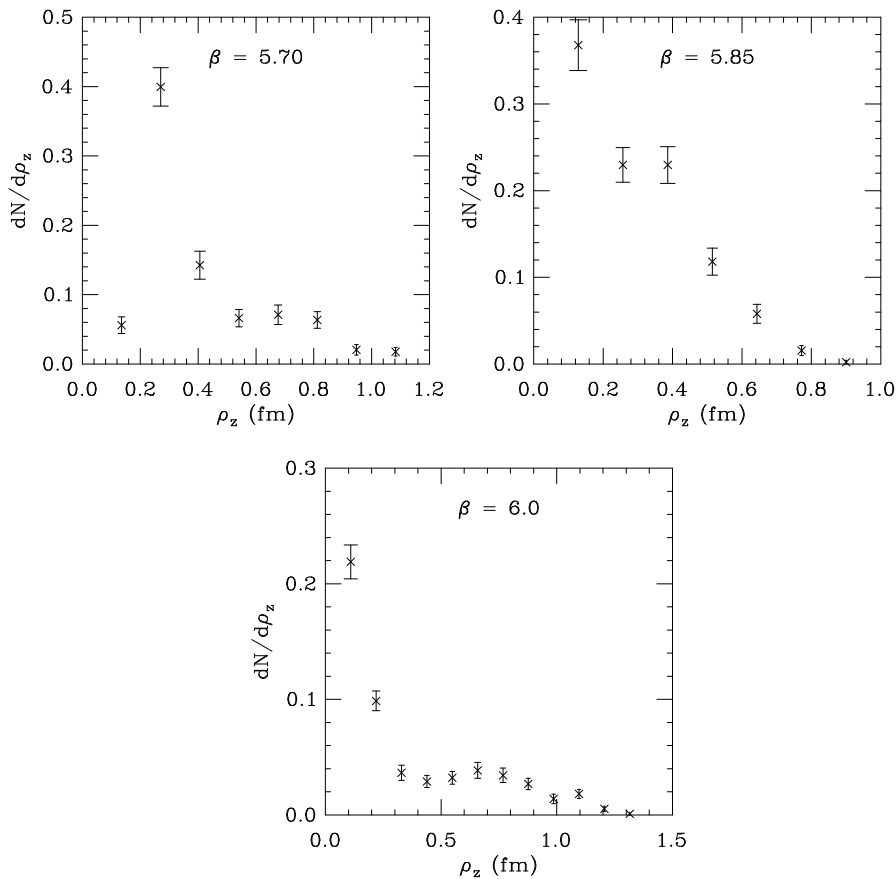}}
\caption{
 Size distribution of zero modes on various SU(3) ensembles.
}
\label{dN_su3}
\end{figure}

The conclusions remain the same for the various SU(3) ensembles that
we studied.
The size distributions are plotted in Figure~\ref{dN_su3}.
The $\beta=5.7,\ 5.85\ {\rm and}\ 6.0$ ensembles come from lattices with linear extent
roughly equal to $1.4$ fm, $1.04$ fm and $1.6$ fm, respectively. 
Clearly the size distribution on the $\beta=5.85$ ensemble suffers
strongly from finite volume effects whereas the $\beta=6.0$ ensemble
is not affected as much.
We should remark that we do not see any evidence for a finite volume
effect in the computation of the topological susceptibility. This is
probably because the size distribution of the individual zero modes
is not that relevant for the global topology which only depends in
principle on the net number of level crossings and not on the size
and shape of these crossing modes.

\section{Discussion}

A probe of pure lattice gauge field ensembles using Wilson fermions
has revealed that the gauge fields are not continuum like on the
lattice at gauge couplings that are typically considered to be
weak. If they were continuum like, we should have seen evidence that
$\rho(0;m)$ is non-zero at a single value of $m$ or in a region in $m$
that is of the order of the lattice spacing. Furthermore, we should
have seen a symmetry in the spectrum at values of $m$ on either side
of the point (or region) where $\rho(0;m)$ is non-zero. Instead, we
found that $\rho(0;m)$ is non-zero in a region $m_1 \le m \le 2$. In
the continuum limit, there is evidence that $m_1$ goes to zero and
that $\rho(0;m)$ goes to zero away from $m=0$. However, the spectral
distribution does not show evidence for a symmetry as 
$m\rightarrow -m$.

A remark on the approach of $\rho(0;m)$ to the thermodynamic limit at
a fixed lattice gauge coupling is in order.
For a small lattice, with linear size of the order of the extent $N_t$
for which the finite temperature deconfinement transition occurs at
the given gauge coupling, the number of very small eigenvalues of
the Wilson-Dirac operator will be essentially zero. When the lattice size
is increased this number will grow rapidly, leading, for a while, to a rapid
increase in the extracted estimate of $\rho(0;m)$ and then leveling
off at the infinite volume value of $\rho(0;m)$. We found that this
happens for a linear size about twice the extent $N_t$ mentioned above.
%A table of $\beta_c(N_\tau)$ can be found
%in~\cite{Fingberg} for SU(3) and SU(2).  For example, one needs to
%have a lattice volume bigger than $8^4$ if one is using an SU(3) gauge
%coupling of $\beta=6.0$ and one needs to work with a lattice larger
%than $14^4$ at $\beta=6.38$.  Only then will one see a value of
%$\rho(0;m)$ close to the thermodynamic limit.

The density of level crossing zero modes, $dN/dm$,
of the Hermitian Wilson--Dirac operator is
in accordance with the behavior of $\rho(0;m)$. In spite of a large
number of levels crossing zero in the bulk of $m_1 \le m \le 2$, we found
that the topological susceptibility is unchanged by these small
localized modes.
We therefore interpret them as due to ultra-violet fluctuations. The size
distribution of the zero modes is dominated by these small modes.
However, the distribution, after we remove these small modes, does not show
any clear peak at a physical scale. Some broad peaks
we see in are explained as a consequence of finite volume effects.
%The gauge field background is rough and one cannot provide any
%input to models which treat gauge fields as being made up of 
%a smooth collection of instantons and anti-instantons.

We finally remark that all our studies in this paper have been
on pure gauge field ensembles. There is a prediction for the
spectral distribution of the Wilson-Dirac operator in full QCD
using a continuum
chiral lagrangian~\cite{Sharpe}. The prediction resembles the
second scenario presented in section II and is qualitatively
different from the result we have obtained for pure gauge ensembles.
It would
be interesting to test this prediction by numerical simulations of
full QCD with Wilson fermions in the supercritical region.
There is, however, a technical problem in using standard
Hybrid Monte Carlo type algorithms for such simulations: the
system will be locked in a single topological sector (with topology
defined as half the difference of negative and positive eigenvalues
of the hermitian Wilson-Dirac operator at the supercritical mass, $m_d$,
where the simulation is carried out).
This is due to the fact that a change
in topology will require a change of net level crossings in the
region $0 < m < m_d$. However, the spectral flow has to be smooth as
we update the configurations using classical dynamics in 
HMC type algorithms. Therefore, at some point in the
change of topology the level crossing would need to occur at $m_d$, 
but such a configuration has a vanishing fermion determinant and hence
can not be reached. Modifications of the HMC algorithm
to circumvent this problems would need to be developed before
a study of the spectral flow in full QCD in the supercritical
region can be attempted.

\acknowledgements
This research was supported by DOE contracts 
DE-FG05-85ER250000 and DE-FG05-96ER40979.
Computations were performed on the CM-2 and QCDSP at SCRI.

\end{document}